\documentclass[a4paper,12pt]{article}
\usepackage{epsfig} 
\usepackage{hyperref} 
\usepackage{amsmath}
\usepackage{psfrag}

\newcommand{\rem}[1]{}

\newcommand{\be}{\begin{equation}}
\newcommand{\ee}{\end{equation}}
\newcommand{\ba}{\begin{eqnarray}}
\newcommand{\ea}{\end{eqnarray}}

\newcommand{\ben}{\begin{displaymath}}
\newcommand{\een}{\end{displaymath}}
\newcommand{\ban}{\begin{eqnarray*}}
\newcommand{\ean}{\end{eqnarray*}}
\newcommand{\brr}{\begin{array}}
\newcommand{\err}{\end{array}}

\newcommand{\bc}{\begin{center}}
\newcommand{\ec}{\end{center}}

\begin{document}

\title{ Systematic calculation of molecular vibrational spectra through a
complete Morse expansion.  }

\author{Andrea Bordoni and Nicola Manini\\
{\it Dipartimento di Fisica, Universit\`a di Milano,}\\
{\it Via Celoria 16, 20133 Milano, Italy}
}

\date{June 19 2006}
\maketitle

\begin{abstract}
We propose an accurate and efficient method to compute vibrational spectra
of molecules,
based on exact diagonalization of an algebraically calculated matrix based
on powers of Morse coordinate.
The present work focuses on the 1D potential of diatomic molecules: as
typical examples, we apply this method to the standard Lennard-Jones
oscillator, and to the {\it ab-initio} potential of the H$_2$ molecule.
Global cm$^{-1}$ accuracy is exhibited through the H$_2$ spectrum, obtained
through the diagonalization of a $30\times 30$ matrix.
This theory is at the root of a new method to obtain globally accurate
vibrational spectral data in the context of the multi-dimensional potential
of polyatomic molecules, at an affordable computational cost.
\end{abstract}


\section{Introduction}

Extremely accurate rovibrational levels are obtained routinely for
polyatomic molecules by modern spectroscopical techniques.
Despite intense effort and significant progress, the accuracy and
reliability of theoretical calculations is still lagging far behind
expecially in the region of high overtones.
Many present-day quantum-chemical calculations of spectroscopical
properties are based on some power expansion of the adiabatic potential
energy surface (PES) at its minimum
\cite{Carter97,Wyatt98,Pochert00,Stuchebrukhov93}.
Accurate treatment of the kinetic and second-order potential term provides
the harmonic normal coordinates and frequencies. Higher-order (cubic,
quartic) power corrections usually account for anharmonic effects.
The ``force field'' parameters are often derived by {\it ab-initio}
methods, or adjusted to experimentally observed levels.
Either way, simple power terms in the normal coordinates fit very poorly
the global shape of the actual adiabatic potential energy surface (PES)
away from its minimum \cite{DelMonte05}.
As a result, methods based on such power expansions retain spectroscopical
accuracy only up to few thousand cm$^{-1}$
\cite{Csaszar97,Koput98,Miani00,Callegari03} in ordinary simple molecules
such as H$_2$O, or H$_2$C$_2$; the accuracy rapidly deteriorates in the
spectral region where dissociation is approached.

To improve this situation, expansions of the PES in suitable internal
coordinates \cite{Jensen00}, for example of the Morse type
\cite{Jensen88,Lemus04,Halonen88}, have been proposed.
In this class of schemes, difficulties arise due to the generation and the
need to diagonalize very non-sparse matrixes, which impose uncontrolled
approximations, involving the neglect of a huge number of matrix elements
\cite{HCAO1}.
To improve this state of affairs
here we propose a method, based on the Morse oscillator, which combines the
accuracy of a theory which includes anharmonicity right from the start with
the simplicity of the harmonic expansion (basis completeness, simple
algebraic rules for the matrix elements, sparse coupling matrixes).
The present work focuses on the one-dimensional (1D) potential,
representing diatomic molecules.
In this context where trivial exact numerical solution of the vibrational
problem is available, the theory is formulated in its full detail, and its
accuracy is tested against the PES of realistic diatomic potentials.
This theory is constructed with careful attention to make it suitable as
the basic building block for the successive extension to the
multi-dimensional potentials of polyatomic molecules, which will be the
subject of future work.

Section \ref{model:sec} introduces the formalism, whose application is then
investigated in Sec.~\ref{Tests}. Conclusions are then drawn in
Sec.~\ref{discussion:sec}.

\section{The model}
\label{model:sec}

The Morse potential was introduced \cite{Morse_1929} with the
purpose of producing an accurate description of typical spectra of diatomic
molecules through an analytically solvable Schr\"{o}dinger problem.
Through standard manipulations of the three dimensional (3D) spherical problem, we come to what is usually named Morse equation
\begin{equation}\label{Ham_Morse}
\hat{H}_M \Psi(x) \equiv
\left[-\frac{\hat{p}_x^2}{2 \mu} + V_M(x) \right] \Psi(x)=
W \Psi(x)
\,,
\end{equation}
where $\mu$ is the reduced molecular mass, and %$V_M(x)\equiv V_0 [e^{-2\alpha (x-x_0)} - 2e^{-\alpha (x-x_0)}]$ 
the {\it Morse Potential} function
\begin{equation}\label{Elem-TermNew}
V_M(x)\equiv V_0 \left\{[v(x)]^2-1\right\},\quad {\rm with}\ 
v(x)=e^{-\alpha (x-x_0)}-1
.
\end{equation}
This potential function is parameterized by the three quantities $V_0$
(depth of the well), $\alpha$ (shape parameter), and $x_0$ (position of the
minimum).
$V_M$ reflects the main features of a diatomic molecular potential, namely
one well defined minimum, a finite dissociation energy for large $x$, a
strongly repulsive region at $x \ll x_0$, a finite number $L$ of bound
states and a continuum of scattering states.
The Morse potential represents an intrinsically anharmonic oscillator, and
indeed it was introduced originally to deal with anharmonicity exactly: it
has the great advantage that the corresponding Schr\"{o}dinger equation
(\ref{Ham_Morse}) is solvable exactly in 1D, and approximately when it
represents the radial equation of a 2-3D problem.

In terms of the dimensionless parameter
\begin{equation}\label{sdef}
s=\frac{\sqrt{2 m V_0}}{\hbar \alpha}-\frac{1}{2}
\,,
\end{equation}
the bound-state eigenfunctions can be written as
\begin{equation}\label{Morse-Wave}
\Psi_n(x)= N_n \, e^{-y(x)/2} \, y(x)^{s-n} \, L_n^{2(s-n)}(y(x))
\,,
\end{equation}
where 
\begin{equation}\label{ydef}
y(x)=(2s+1)\, e^{-\alpha(x-x_0)}
.
\end{equation}
$L_n^{\mu}(z)=\frac{e^{z} z^{-\mu}}{n!} \frac{d^n}{dz^n} (z^{\mu+n} e^{z})$
is a generalized Laguerre polynomial, and
\begin{equation}\label{Nn}
N_n=\alpha^{1/2}\,\left[\frac{\Gamma(2s-n+1)}{2(s-n)\Gamma(n+1)}\right]^{\frac{1}{2}}\,
\end{equation}
is a normalization constant.
For integer $n=0,1,2,\dots [s]$
($[s]$ indicates the largest integer $\leq s$), the square-integrable 
wavefunctions $\Psi_n(x)$ represent the $[s]+1$
bound states, for the specific physical parameters of the system.
The corresponding energy eigenvalues are
\begin{equation}\label{Morse-Ener}%Provvisoria
W_n(x)= -\frac{\alpha^2\hbar^2}{2 \mu}(s-n)^2.
\end{equation}

In addition to its many advantages, the Morse potential has a few
drawbacks: mainly, it does not approach the dissociation limit (here $E=0$)
with a power law as realistically expected, and it does not quite diverge
at $x=0$.
On the other hand, a more realistically accurate potential $V(x)$ can be
obtained by adding a suitable correction $V_d(x)\equiv V(x)-V_M(x)$ to the
starting Morse potential.
This correction is often written as an expansion, whose detailed form is
determined according to some convenient recipe.
Several approaches in this direction are documented in the literature, from
the perturbed Morse oscillator model (PMO)
\cite{Huffak_1976_1,Huffak_1982,Makar_1990}, to models based on algebraic
methods \cite{Iac-Lev_book,VanRoos-PhD,Oss-rev}.
The PMO approach is based on an expansion of $V_d(x)$ in a power series of
$v(x)$
\begin{equation}\label{VPMO}
V_d(x) = \sum_{i=4} a_i \, \left[v(x)\right]^i
,
\end{equation}
then treated perturbatively.
In contrast, algebraic approaches based on dynamical symmetries lead to an
expansion of the Casimir operator of a Lie group, representing the whole
Morse Hamiltonian \cite{Iac-Lev_book,VanRoos-PhD,Oss-rev,Iac-Lev_I}.
Unfortunately, as this operator is of mixed kinetic and potential
character, it is hard to relate this second type of expansion to a well
defined molecular problem.

Here we propose a substantial improvement of the PMO.
This method is based on an expansion of the form
\begin{equation}\label{Morse-Exp}
V_d(x)= \sum_{i=3}^{N_{max}} a_i \, \left[v(x)\right]^i
,
\end{equation}
including powers of $v(x)$ up to order $N_{max}$, usually a small integer,
e.g.\ 4 or 10.
The differences with the PMO are two:
(i) the inclusion of the $[v(x)]^3$ term, which the PMO approach must omit
due to the need to limit the number of low-order terms contributing to
that perturbative scheme, and which allows much better flexibility and
accuracy, as shown is Sec.~\ref{Tests};
(ii) the exact, rather than perturbative solution of the resulting quantum
mechanical problem.
To this respect, the implementation of algebraic rules for the calculation
of the matrix elements allows us to treat the third-order term on the same
footing as all other terms in $V_d$: we compute their matrix elements, and
subsequently diagonalize a sparse matrix.

The choice (\ref{Morse-Exp}) of the form of the potential correction
realizes a satisfactory compromise between the quest of a locally accurate
parameterization of the target potential $V(x)$ by means of a globally
meaningful expansion, and the need of analytical expressions of its matrix
elements.
In particular, the expansion (\ref{Morse-Exp}) satisfies the minimal
requirement of being lower-bounded everywhere, provided that the last
coefficient $a_{N_{max}}$ is positive.
This feature holds independently of $N_{max}$, in contrast to the
traditional expansion in powers of $x$ where the truncation to odd
$N_{max}$ leads to lower-unbounded potentials regardless of the sign of the
coefficients \cite{DelMonte05}.
We will show in specific applications that the chosen form
(\ref{Morse-Exp}) is sufficiently flexible to approximate a realistic
molecular potential $V(x)$ substantially better than $V_M(x)$, and, as long
as $V(x)$ is not ill behaved, in principle arbitrarily well on a very wide
energy range from the bottom of the well to dissociation.
The Morse parameters and the coefficients $a_i$ are determined by fitting
the computed adiabatic potential $V(x)$, thus leading to a fully {\it
ab-initio} method to compute the molecular vibrational spectra, as
demonstrated in a few examples in Sec.~\ref{Tests}.
Once the parameters are determined, by taking advantage of the algebraic
properties of $V_d(x)$, in Sec.~\ref{matel_morse:sec} we build the explicit 
matrix representation of the Hamiltonian
\begin{equation}\label{MorsePertSint}
H= H_{M} + V_d(x)
\end{equation}
on a variational basis.  We then perform a full diagonalization, thus
obtaining eigenvalues and eigenfunctions for this quantum mechanical
problem.
In Sec.~\ref{matel_morse:sec} we consider the basis of Morse eigenstates
(\ref{Morse-Wave}).
In Secs.~\ref{beyond:sec} and \ref{matel_quasi:sec}, we introduce an
extended basis, capable to include states in the continuum, thus improving
the variational convergency of the vibrational states close to molecular
dissociation.

\subsection{Matrix elements on Morse basis}
\label{matel_morse:sec}

The expression of the matrix elements of an operator of the form
$[e^{-\alpha (\hat{x}-x_0)}]^t$, such as appears in power terms
$[\hat{v}(x)]^i=[e^{-\alpha (\hat{x}-x_0)}-1]^i$, Eq.~(\ref{Morse-Exp}), on the
Morse basis (\ref{Morse-Wave}) are available in the literature
\cite{Sage1}.
For convenience we report here the analytical expression for these matrix
elements  computed by exploiting basic properties of Laguerre
polynomials:
\begin{eqnarray}\label{MorseMatEl}
&&\langle n | [e^{-\alpha (\hat{x}-x_0)}]^t | m \rangle =\frac{(-1)^{n + m}}{(2s+1)^t} N_n N_m \Gamma[2s - m - n + t]\times\\ \nonumber
   && \sum_{\ell=0}^{\min(n,m)}\binom{n - m - 1  + t}{ n - \ell} 
     \binom{m - n -1 + t}{ m - \ell} 
     \binom{2s - 1 - m - n + \ell + t}{ \ell}\,,
\end{eqnarray}
where $N_n,N_m$ are the normalization constants of
Eq.~(\ref{Nn})\footnote{
  Equation (\ref{MorseMatEl}) holds for real $t \geq 0$, but we are
  interested only in integer $t$: in this case, the first non vanishing
  term occurs for $\ell=(1-\delta_{n,m}) \max(0,\,\min(n,m)+1-t)$.
}.
The matrix elements of a term $\hat{v}(x)^i$, are
obtained by combining expressions of this kind.
Note that the matrix element (\ref{MorseMatEl}) is non vanishing for any
$n,m$: therefore the resulting matrix is not sparse.
Equation (\ref{MorseMatEl}) permits to construct the matrix representation
of the Hamiltonian (\ref{MorsePertSint}), which is exact within the space
of Morse bound states.
The molecular problem is thus reduced to the diagonalization of a
$([s]+1)\times([s]+1)$ matrix: clearly it yields exact eigenvalues
and eigenvectors in the trivial limit $H \to H_M$ (i.e.\ $V_d(x)=0$).

When $V_d(x)\neq 0$,  $H$ mixes states within the Morse basis to states
orthogonal to it.  The matrix representation of $H$ is therefore only
approximate, and provides a variational extimate of the energy of the lowest
bound states.
Due to the imperfect overlap of the subspace of Morse bound states to the
subspace of exact bound states of $H$, we expect that when the correction
$V_d(x)$ to the Morse problem becomes important in
Eq.~(\ref{MorsePertSint}), the accuracy of the description deteriorates.
This point is studied in detail in the next section.

\subsection{The incompleteness of the Morse basis}
\label{incompleteness:sec}

Morse potential bound states are clearly incomplete for generating the full
infinite-dimensional Hilbert space: the un-normalizable wavefunctions of
the continuum should also be included to complete it.
We expect that the incomplete Morse basis should provide a good description
of the actual spectrum for potentials weakly deformed from the Morse shape
(small coefficients $|a_{i}|\ll V_0$), especially for the lowest bound
states.  For substantial distortion from the Morse shape, the incomplete
basis could fail badly, especially for states close to dissociation, and
even for intermediate levels.

\begin{figure}
\centerline{\epsfig{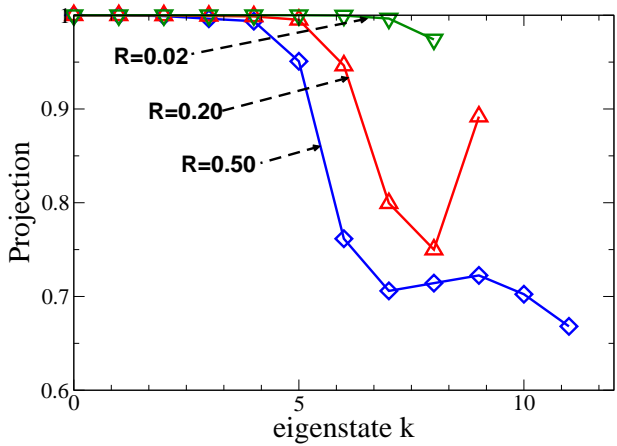}}
\caption{
The projection $\sum_{j=0}^{[s]} |\langle\psi_j|\varphi_k\rangle|^2$ of the
$k^{\rm th}$ eigenstate on the subspace spanned by the Morse basis, for
increasing relative importance $R$ of the deformation $V_d(x)$ added to a
Morse potential characterized by $[s]+1=9$ bound states
($\hbar=\mu=1$, $s=8.34$).
\label{FigProjMorse}}
\end{figure}

The incompleteness problem is illustrated quantitatively in
Fig.~\ref{FigProjMorse}, which reports the total projection
$\sum_{j=0}^{[s]} |\langle\psi_j|\varphi_k\rangle|^2$ of the
finite-differences numerical eigensolutions $\varphi_k(x)$ of the
Schr\"{o}dinger equation associated to the potential $V(x) = V_M(x)+ a_4 \,
[v(x)]^4$ to the Morse basis (\ref{Morse-Wave}).
As expected, the projection of the exact eigenstates of $H$ on the subspace
spanned by the Morse basis (\ref{Morse-Wave}) gradually reduces from unity
(especially for the highest excited states) as the relative weight
\begin{equation}\label{RelWeight}
R=\frac{1}{V_0}  \sum_{i=3}^{N_{max}} |a_i|
\end{equation}
(here $R=a_4/V_0$) of $V_d(x)$ increases.
The increase in well depth caused by increasing $R$ produces an increase in
the number of bound states from $9$ (pure Morse), to $12$, for $R=0.5$, and
this makes the problem of overlaps even more severe.
Note that the lowest eigenvectors retain a projection near unity
even for large $R$.
This indicates that the lowest states are determined fairly accurately even
within the substantially incomplete basis of a strongly deformed Morse
oscillator.
In contrast, the accuracy of the states close to dissociation 
deteriorates rapidly.

\begin{figure}
\psfrag{2P}{2\%}
\centerline{\epsfig{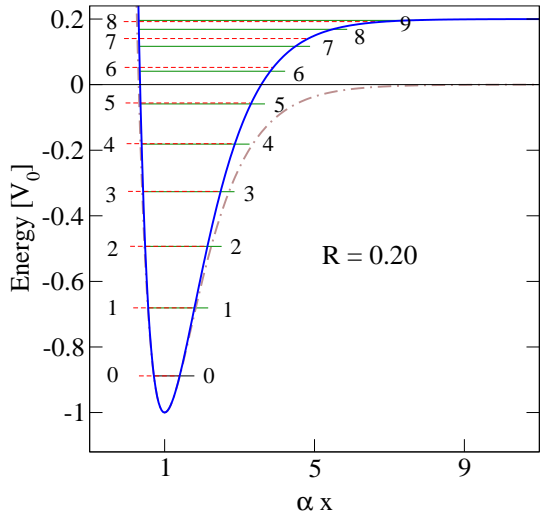}}
\caption{\label{FigMorseR20}
The Schr\"{o}dinger bound states of the anharmonic oscillator of potential
energy $V(x)=V_M(x) + [v(x)]^{4}$ (solid curve).
Comparison between exact eigenenergies $E^{ex}_k$ (solid levels) and
eigenvalues $E^{alg}_k$ obtained by diagonalization of the $H$ matrix
within the Morse incomplete subspace (dashed levels).
The good accuracy of the lowest levels deteriorates for those close to
dissociation, and in particular the highest bound state $n=9$ is obviously
missing. The pure Morse well $V_M(x)$ is also outlined for comparison
(dot-dashed curve).
}
\end{figure}

Figure \ref{FigMorseR20} illustrates the effect of basis incompleteness on
the approximate eigenenergies $E^{alg}_k$ obtained by numerical
diagonalization of the matrix representation of $H$, obtained by applying
Eqs.~(\ref{Morse-Ener}) and (\ref{MorseMatEl}).
As the overlap analysis suggests, low levels remain accurate, while close to
dissociation the method relying on the Morse basis gives very poor results.

The above calculations focus on the case $V_d(x)\propto [v(x)]^4$.
A similar analysis can be carried out for any term $[\hat v(x)]^i$ or linear
combination thereof, with the same definition (\ref{RelWeight}) for $R$ and
qualitatively similar results.

\subsection{Beyond the Morse basis}
\label{beyond:sec}

To improve the accuracy of the exact diagonalization method close to
dissociation, it is necessary to apply it on a basis which can be
systematically enlarged virtually to cover the entire Hilbert space.
In recent years, much work has been done to treat at least in an
approximate manner the continuum part of the spectrum, mostly in the
context of molecules. The main reason is that energy regions near
dissociation are becoming available experimentally: in such regions the
interaction between the highest-energy bound states and the lowest-energy
section of the continuum spectrum is supposed crucial.
The quest for manageable approximations of the continuum spectrum has
followed different routes, including analytical methods (wavefunctions 
manipulation in coordinate representation \cite{PerBer1,PerBer2}), to
essentially algebraic treatments related to the dynamical-symmetry
viewpoint \cite{su11} or of supersymmetric (SUSY) quantum mechanics derivation
\cite{SusyRev}.

An algebraic approach is the most suited for a subsequent generalization to
 potentials of the kind of Eq.~(\ref{Morse-Exp}). Moreover, if an algebraic form
of the dynamical variables is available, the computation of matrix elements
and other interesting quantities can be done by exploiting simple algebraic
relations.
We sketch here the main ideas followed in the literature, with particular
attention to what is useful for our specific work.

The basic idea is to build appropriate generalized ladder operators connecting the different wavefunctions of a given family, and subsequently express all relevant quantities in terms of these operators. The ladder operators on one side allow the identification of the dynamical symmetry of the physical model under consideration, and on the other side lead to an algebraic expression for all relevant dynamical variable, including the Hamiltonian. 
A similar construction was made for the Morse potential \cite{DongLemus},
starting from the Morse bound-state eigenfunctions (\ref{Morse-Wave}). 

Eventually, we need a basis where suitable ladder operators act: this basis
must be complete in $\mathcal{L}^2[(0,\infty),dy/y]$.
To this purpose we introduce the following family of
wavefunctions \cite{su11}
\begin{equation}\label{QNSB}
\phi_n^{\sigma}(y)=\sqrt{\frac{\alpha n!}{\Gamma (2\sigma +n)}}
\,y^{\sigma} e^{-\frac{y}{2}}L_n^{2\sigma -1}(y),~ n=0,~1,~2,\dots
\end{equation}
labeled by a real parameter $\sigma >0$.
Following Ref.~\cite{Ben-Mol-Al}, we refer to $|\phi_n^{\sigma}\rangle$ as
{\it quasi number state}.
Formally, these functions resemble Morse bound states
(\ref{Morse-Wave}). However they are substantially different.
In particular for any $\sigma>0$, the quasi number states do form a
complete orthonormal set in $\mathcal{L}^2[(0,\infty),dy/y]$ \cite{su11}.
This same set was introduced in Ref.~\cite{Ben-Mol-Al}, on the basis of
considerations related to SUSY quantum mechanics, and applied to the
context of molecular spectra in Ref.~\cite{Tennyson82,Tennyson04}.

The construction of generalized ladder operators realized for Morse bound
states can be repeated starting from Eq.~(\ref{QNSB}) \cite{su11}. The
three resulting operators are
\begin{eqnarray}
\nonumber
\hat{K}_{-} &=&
 (\sigma \hat{I}+ \hat{n}) -\frac{\hat{y}}{2} -y\frac{d}{dy}
\quad\left(=(\sigma \hat{I}+ \hat{n}) -\frac{\hat{y}}{2}  +\frac{i}{\hbar \alpha}\hat{p}_x \right)\\
\label{Kdef}
\hat{K}_{+} &=&
 (\sigma \hat{I}+ \hat{n}) -\frac{\hat{y}}{2} +y\frac{d}{dy}
\quad\left(= (\sigma \hat{I}+ \hat{n}) -\frac{\hat{y}}{2} -\frac{i}{\hbar \alpha}\hat{p}_x\right)\\ 
\nonumber
\hat{K}_{0} &=& \sigma \hat{I}+ \hat{n}
\,,
\end{eqnarray}
where $\hat{n} \phi_n^{\sigma} = n \phi_n^{\sigma}$.
$\hat{K}_{-}$, $\hat{K}_{+}$ and $\hat{K}_{0}$ act as generalized step up,
step down and number operator repeatedly on the states represented by
Eq.~(\ref{QNSB}):
\begin{eqnarray}
\hat{K}_{-} |\phi_n^{\sigma}\rangle  &=& C_{n} |\phi_{n-1}^{\sigma }\rangle\\
\hat{K}_{+} |\phi_n^{\sigma}\rangle  &=& C_{n+1} |\phi_{n+1}^{\sigma} \rangle\\
\hat{K}_{0} |\phi_n^{\sigma}\rangle  &=& (\sigma + n) |\phi_n^{\sigma}\rangle 
\,,
\end{eqnarray}
where 
\begin{equation}
C_{n}=\sqrt{n(n+2\sigma-1)}
\,.
\end{equation}
The definition (\ref{Kdef}) can be inverted, to obtain the operatorial form
of momentum and Morse coordinate
\begin{eqnarray}\label{pxK}
\hat{p}_x &=& \frac{i \hbar \alpha}{2} (\hat{K}_{+}-\hat{K}_{-})\\ 
\hat{y} &=& 2 \hat{K}_{0} - (\hat{K}_{+}+\hat{K}_{-})\,. \label{yK}
\end{eqnarray}
Accordingly, the whole Hamiltonian (\ref{Ham_Morse}) may be expressed in algebraic form, as done in Ref.~\cite{su11}: the matrix elements $\langle \phi_m^{\sigma}|\hat{H}_M| \phi_n^{\sigma}\rangle$ can be computed algebraically.
The only non-vanishing matrix elements are those for which $m=n,~n \pm 1$: the Morse Hamiltonian matrix is tridiagonal on the basis $|\phi_{n}^{\sigma } \rangle$.
All the above properties hold regardless of the value of $\sigma$, for which 
a convenient choice will be made in Sec.~\ref{matel_quasi:sec}.

An instructive similarity to the algebraic structure of the harmonic oscillator
may be obtained by following the SUSY approach:
in SUSY quantum mechanics the Morse potential satisfies the
condition of being {\it shape-invariant}, so that its eigenvalues and
eigenvectors can be obtained in algebraic form. This leads to the
construction of Morse generalized ladder operators of the form
\begin{eqnarray}\label{SUSYGenLadd}
\hat{A}(q)&=&
q\hat{I} - \frac{\hat{y}}{2} + \frac{i}{\hbar \alpha}\hat{p}_x\\ \nonumber
\hat{A}^{\dagger}(q)&=&
q\hat{I} - \frac{\hat{y}}{2}  - \frac{i}{\hbar \alpha}\hat{p}_x \ ,
\end{eqnarray}
parameterized by the real quantity $q$.
The Morse Hamiltonian (\ref{Ham_Morse}) can be expressed in factorized form
\begin{equation}\label{HMAlg}
\hat{H}_M=\frac{\hbar^2 \alpha^2}{2\mu} [\hat{A}^{\dagger}(s)\hat{A}(s) -s^2]\ ,
\end{equation}
in terms of the operators of Eq.~(\ref{SUSYGenLadd}) with $q=s$, and the 
formal analogy with harmonic oscillator algebra is evident.
Define $|\Psi_0^q \rangle$ as the only state annihilated by $\hat{A}(q)$: 
$\hat{A}(q)|\Psi_0^q \rangle=0$ \cite{Ben-Mol-Al}. 
For $q=s$, $|\Psi_0^q \rangle$ coincides with the ground state of Morse 
Hamiltonian, Eq.~(\ref{Morse-Wave}).
The other excited states can be obtained algebraically as
\begin{equation}\label{PhiMorseGen}
|\Psi_n^s \rangle = F_n \prod_{\ell=1}^{n} \hat{A}^{\dagger}(s -n +\ell)|\Psi_0^{s-n}\rangle\ ,
\end{equation}
where $F_n$ are normalization constants.
Equation (\ref{PhiMorseGen}) is reminescent of the ladder representation of 
the harmonic oscillator with two important differences: (i) the excited state 
$ 
|\Psi_n^s \rangle$ is generated starting from the ground state of a 
{\it different but related} Morse problem;
(ii) the ladder operators involved depends explicitly on the step number.

Due to the particularly simple dependence of the operators 
$\hat{A},~\hat{A}^{\dagger}$ on the parameter $q$, different choices of the 
step number $q$ dependency generate new  algebraic
families of wavefunctions, which do not coincide with Morse eigenstates, 
but have similar algebraic properties. In particular, setting $q=\sigma+n$, 
starting from the ket  $|\phi_0^{\sigma} \rangle\equiv |\Psi^s_0\rangle$ 
annihilated by $\hat{A}(\sigma)$, the following family of wavefunctions 
is generated:
\begin{equation}\label{PhiGen}
|\phi_n^{\sigma} \rangle = G_n^{\sigma}\prod_{\ell=1}^{n} \hat{A}^{\dagger}(\sigma -1 + \ell) |\phi_0^{\sigma} \rangle \ ,
\end{equation}
where $G_n^{\sigma}$ are appropriate normalization constants.
By comparing Eq.~(\ref{SUSYGenLadd})  with Eq.~(\ref{Kdef}), 
the following relations emerge:
\begin{eqnarray}
\hat{A}(\sigma+n) \phi_n^{\sigma} &=&  \hat{K}_{-} \phi_n^{\sigma}\\ \nonumber
\hat{A}^{\dagger}(\sigma+n) \phi_n^{\sigma} &=&  \hat{K}_{+} \phi_n^{\sigma} \ .
\end{eqnarray}
This means that, for this special ($n$-dependent) choice $q=\sigma+n$, 
the $\hat{A},~\hat{A}^{\dagger}$ operators assume the structure of the family 
of ladder operators (\ref{SUSYGenLadd}), and the states (\ref{PhiGen}) 
coincide with those of Eq.~(\ref{QNSB}).

These SUSY considerations allow us to see the complete basis of Eq. 
(\ref{QNSB}) as a natural algebraic modification of the Morse basis, 
with the nontrivial advantage that all matrix element of operators expressed 
as functions of  $\hat{A}(q)$ and $\hat{A}^{\dagger}(q)$ (see Eqs.~(\ref{pxK}) and (\ref{yK})) can be computed by purely algebraic means.
We will  take advantage of this feature in the next section to evaluate the 
matrix elements of all operators involved in the expansion (\ref{Morse-Exp}).

\subsection{Matrix elements on the quasi-number basis}
\label{matel_quasi:sec}

In analogy to Sec.~\ref{matel_morse:sec}, we derive algebraic matrix elements 
of powers of the operators $e^{-\alpha(\hat{x}-x_0)}$, which appear in the 
expansion (\ref{Morse-Exp}) of $V_d(x)$, on the basis (\ref{QNSB}) of 
quasi number states.
In terms of the operators $A(q)$, $A^{\dagger}(q)$ defined in
Eq.~(\ref{SUSYGenLadd}),
the Morse variable $\hat{y}$ and the momentum operator can be written
as:
\begin{eqnarray}\label{YYdef}
\hat{y}=(2s+1) e^{-\alpha(\hat{x}-x_0)} &=& 2s\hat{I} - \left[\hat{A}^{\dagger}(s)+\hat{A}(s)\right]
\\
\hat{p}_x &=& \frac{\hbar \alpha}{2i} \left[\hat{A}(s)-\hat{A}^{\dagger}(s)\right] \,.\label{PXdef}
\end{eqnarray}

The  $\hat{A}(s), ~\hat{A}^{\dagger}(s)$ matrix elements are computed explicitly by 
observing that $\hat{A}(s)=\hat{A}(\sigma+n) + (s - \sigma -n)\hat{I}$
(and analogously for $\hat{A}^{\dagger}$):
\begin{eqnarray}\label{MatelA}
\langle \phi_m^{\sigma} | \hat{A}(s)| \phi_n^{\sigma} \rangle &=& C_n \delta_{m,n-1} + (s - \sigma -n) \delta_{m,n}\\ \nonumber
\langle \phi_m^{\sigma} | \hat{A}^{\dagger}(s)| \phi_n^{\sigma} \rangle &=& C_{n+1} \delta_{m,n+1} + (s - \sigma -n) \delta_{m,n} \ .
\end{eqnarray}

Based on the algebraic relations above, it is straightforward to obtain the
matrix elements of all the relevant operators. In particular, the matrix
elements of $\hat{H}_M$ are
\begin{eqnarray}
\langle \phi_m^{\sigma}|\hat{H}_M|\phi_n^{\sigma} \rangle &=& 
\hbar \alpha \left(\frac{2 V_0}{\mu}\right)^{\frac{1}{2}} \frac{1}{(2s+1)} 
\big\{[C_m^2 - s^2 + (m-s + \sigma)^2]\delta_{m,n} 
\nonumber\\ \label{MatEn}
&+&
(s - \sigma -n)C_m \delta_{m,n+1} +(s - \sigma -m)C_n\delta_{m+1,n}\big\}
\,.
\end{eqnarray}
On the quasi number states basis (QNSB) the Hamiltonian $\hat{H}_M$ is 
therefore tridiagonal. In practice, this is a minor drawback with respect to 
the basis (\ref{Morse-Wave}) of energy eigenstates, since tridiagonal matrices 
are diagonalized extremely quickly.

We are now ready to choose the value of the $\sigma$ parameter: from 
Eq.~(\ref{MatEn}) it is evident that if $\sigma-s$ equals a positive integer 
$d$,
then the off-diagonal matrix element $\langle d|\hat{H}_M |d+1 \rangle$
vanishes: $\sigma$ determines the splitting of the Hamiltonian matrix into two 
blocks. It is natural to set the integer $d$ to the number $[s]$ of Morse 
bound states, i.e.\ 
\begin{equation}\label{SigFix}
\sigma=s-[s] \,,
\end{equation}
 so that $0 < \sigma \leq 1$.
Under this assumption the Hilbert space $\mathcal{H}$ decomposes into the direct 
sum of  a $([s]+1)$-dimensional $\mathcal{H}^{-}$ space (bound states 
(\ref{Morse-Wave})), and an infinite dimensional $\mathcal{H}^{+}$ space 
(Morse continuum spectrum) \cite{Ben-Mol-Al,su11}:
\begin{equation}
\mathcal{H}=\mathcal{H}^{-} \oplus \mathcal{H}^{+}
\end{equation}
\rem{
The $\mathcal{H}^{-}$ space coincides with the subspace generated by Morse bound states (\ref{Morse-Wave}) \cite{Ben-Mol-Al,su11}. 

The basis (\ref{QNSB}) is a complete orthonormal basis composed of functions not much  different from the true bound state eigenfunctions, which allows to take into account the continuum part of the spectrum systematically.}
In all the following we stick to the assumption (\ref{SigFix}).

In the same way of Eq.~(\ref{MatEn}), the matrix elements of  $e^{-\alpha(\hat{x}-x_0)}$ are obtained 
from Eq.~(\ref{YYdef}):
\begin{equation}\label{MatelExp}
\langle \phi_m^{\sigma} | e^{-\alpha(\hat{x}-x_0)}|\phi_n^{\sigma} \rangle= \frac{- C_n \delta_{m,n-1} - C_m \delta_{m,n+1} + 2 (s-[s]+n)\delta_{m,n}}{2s+1}\,.
\end{equation}
The matrix of $e^{-\alpha(\hat{x}-x_0)}$ is also {\it tridiagonal}.
Based on Eq.~(\ref{MatelExp}), all matrix elements of every term in the
expansion (\ref{Morse-Exp}) can be computed exactly.
It is found in particular that the matrix representing $[\hat{v}(x)]^i$ is $(2
i+1)$-band diagonal.\footnote{
Like on the Morse basis, also on QNSB matrix elements of $v(x)^t$ for real
$t$ may be obtained analytically in close form
}
For the reader's convenience, explicit expressions of the nonzero matrix
elements of the $\langle \phi_n|[\hat{v}(x)]^i| \phi_{m}\rangle$ for $i=3$ to 6
are collected in Appendix~\ref{Matel6expl}.

\begin{figure}
\centerline{\epsfig{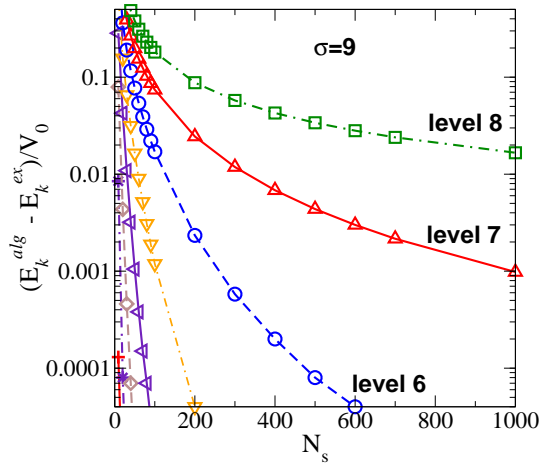}} 
\caption{\label{Tennyson:fig} 
The QNSB with $\sigma=9$, as proposed in Ref.~\cite{Tennyson82,Tennyson04}:
convergence of the bound eigenenergies of the Morse oscillator of
Fig.~\ref{FigProjMorse} as the size $N_s$ of the QNSB (\ref{QNSB}) increases.
This calculation shows that the choice of a value of $\sigma$
substantially larger than our choice Eq.~(\ref{SigFix}) produces a
very slow convergence of the eigenvalues.
}
\end{figure}

Equation~(\ref{QNSB}) represents also the basis of Morse-oscillator-like
functions of Ref.~\cite{Tennyson82,Tennyson04}, but with an initial choice
$\sigma=\frac 12 [2s] +1$.
The value of $\sigma$ changes as $s$ (taken as a free parameter) is
adjusted, but it is generally much larger than the choice of
Eq.~(\ref{SigFix}) considered for the QNSB of the present work and
yielding $\sigma<1$.
As a consequence, the complete set of wavefunctions of
Ref.~\cite{Tennyson82,Tennyson04} decay much faster in approaching
dissociation ($y\to 0$).
This makes convergence of the bound states much slower in that
approach, as illustrated in Fig.~\ref{Tennyson:fig} for the pure Morse
case, where our choice Eq.~(\ref{SigFix}) only requires $N_s=9$ states
to reach full convergence.
Similarly slow convergence is observed for the distorted potential of
Fig.~\ref{FigMorseR20}.
In addition, the matrix elements at the left hand of
Eq.~(\ref{MatelExp}) have generally nonzero value for any $n$, $m$,
thus leading to the computational overhead of nonsparse matrices.

In summary, the following extremely useful properties characterize the 
QNSB (\ref{QNSB}) with the choice $\sigma=s-[s]$:
(i) the first $[s]+1$ states
generate the same Hilbert subspace as the Morse basis of bound states;
(ii) more states can be included simply, with a systematic improvement of
the basis completeness;
(iii) if extended to all $n\geq 0$, it becomes to a complete orthonormal
basis of $\mathcal{L}^2[(0,\infty),dy/y]$, allowing to expand the bound
solutions of an arbitrary 1-dimensional problem, with systematically improvable
accuracy;
(iv) most importantly, it allows to express the matrix elements of all
relevant operators in an algebraic form.
In particular, the matrix relative to a term $[\hat{v}(x)]^{i}$ is $(2 i+1)$-band
diagonal, so that the finite expansion $V_d$ and thus $\hat{H}$ is represented 
by an extremely sparse matrix, which is even a band matrix!
We propose therefore the QNSB as a convenient complete
orthonormal system to study the general problem of a molecular oscillator,
in a way similar to what is usually done in harmonic expansion methods,
but with the advantage of including anharmonicity from the beginning.

\section{Applications}\label{Tests}

\subsection{Convergency with basis extension}

Initially we verify that indeed the QNSB relizes basis completion as
expected.
We start considering the potential of Eq.~(\ref{Morse-Exp}), defined by
\begin{equation}\label{MorsiniTest}
V_d(x)= R V_0 [v(x)]^{4}\, 
\end{equation}
for $R = 0.2$, as in Fig.~\ref{FigMorseR20}.
The eigenvalues $E^{alg}_k$ of the $H$-matrix within the space generated by
the first $N_s$ states in the QNSB converge rapidly to the exact levels
$E^{ex}_k$, those obtained by the numerical finite-difference
integration, represented by solid lines in Fig.~\ref{FigMorseR20}.
The algebraic levels are so accurate that it would be impossible to tell
them apart from the solid lines of Fig.~\ref{FigMorseR20}.
As a quantitative measure of the error of the algebraic method, 
we use a worst-case estimator
\begin{equation}\label{Estimator}
 \Delta_{N_s}:=\frac1{V_0} \max_k\,|E^{alg}_k - E^{ex}_k|\,,
\end{equation}
which equals $\Delta_9 = 2.35\%$ for the calculation based on the Morse basis
illustrated in Fig.~\ref{FigMorseR20}.
With at least $N_s=11$ states included in the QNSB, the bound states reach
the correct number of 9 and the $\Delta_{N_s}$ diminishes fairly rapidly
with increasing basis size: $\Delta_{20}=0.051\%$, $\Delta_{30}=0.028\%$,
$\Delta_{40}=0.019\%$.
It is possible to obtain a good accuracy of
this method even for large values of $R$, namely for potentials
significantly distorted from the Morse shape.
We check that the same form (\ref{MorsiniTest}) of the potential for
$R=1$ yields the correct 14 states with an excellent $\Delta_{30}=0.013\%$.

\begin{figure}
\centerline{\epsfig{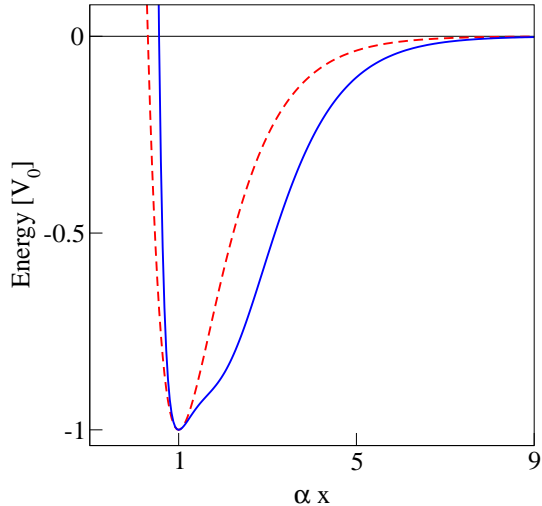}}\caption{Morse potential (dashed) and potential expansion with $V_d(x)= 2.0 V_0 \sum_{i=3}^{6} [v(x)]^{i}$ (solid).\label{Morse-3+4-5+6Peso2}}
\end{figure}

Good convergence against basis completion was verified also in the presence
of multiple power terms, with similar results.
For example, for $V(x)= V_M(x) + 2.0 V_0 \sum_{i=3}^{6} [v(x)]^{i}$,
strongly distorted from the Morse shape (see
Fig.~\ref{Morse-3+4-5+6Peso2}), the $11$ bound states are also obtained
with good accuracy ($\Delta_{30}=0.11\%$, $\Delta_{40}=0.011\%$).

Other tests indicate that a $\Delta_{N_s}$ within a small fraction of percent
is normally realized by including a number $N_s$ of states about twice the
number of bound states.
When high powers of $v(x)$ are included in $V_d(x)$ with substantial
weight, a larger $N_s$ could be required for the same accuracy.
Substantially fewer states can be included if only few low-lying levels are
addressed, and on the contrary, $N_s$ must be substantially increased to
reach convergence of the topmost bound state whenever it happens to occur
very close to the continuum, thus associated to a very extended
wavefunction.

We have therefore established that the method based on the QNSB
(\ref{QNSB}) is reliable and efficient: given an expansion of the actual
molecular potential in the form (\ref{Morse-Exp}), this method provides
accurate eigenvalues and eigenstates, regardless of the relative amount of
deviation from the pure Morse potential.
We come now to the investigation of the flexibility of the expansion
(\ref{Morse-Exp}) to describe realistic potentials.
We consider the Lennard-Jones (LJ) potential, and the adiabatic
potential of the H$_2$ molecule.

\subsection{The Lennard-Jones potential}

We probe the ability of the Morse expansion (\ref{Morse-Exp}) to describe
accurately the standard $(12,6)$ LJ potential
\begin{equation}\label{LJFuncdef}
V_{LJ}(x)=
A_{LJ}\left[\left(\frac{\Sigma}{x}\right)^{12} -
  		\left(\frac{\Sigma}{x}\right)^6\right]
.
\end{equation}
Like the Morse potential, $V_{LJ}$ is a model potential, which however
reproduces well the large-$x$ behavior of the actual interatomic adiabatic
potential of, e.g., a dimer of noble gas.
Due to its power-law dependecies, $V_{LJ}$ can be considered as an
especially ``tough'' potential to approximate by a series of Morse-like
terms.

As the functional form of the target function $V(x)$ is assigned, it is
straightforward to extract a few exact properties, such as well depth 
and derivatives at the minimum, which can be imposed to the 
approximating potential $V_M+V_d$.
Exact equality of these properties can be used to fix a number of
constraints among the $N_{max}+1$ parameters which determine $V_M+V_d$.
The choice of the imposed exact equalities leaves some degree of
arbitrariness to the fitting scheme.
It is possible to enhance global versus local accuracy, by privileging
properties such as the well depth, or rather a number of exact derivatives
at the minimum.
An anlysis of this sort was applied in a different related context 
(Ref.~\cite{Huffak_1976_1}). 

\begin{figure} 
\centerline{\epsfig{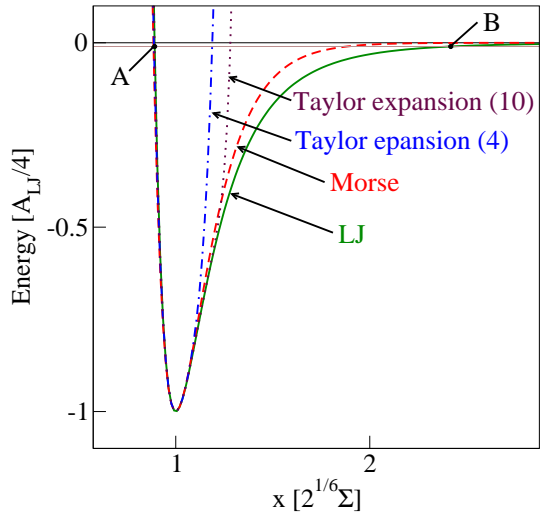}}
\caption{\label{LJMorsebuca}
Approximations of the LJ potential (solid): the Morse potential (dashed),
with equal position of the minimum, second derivative in the minimum, and
well depth; the $4^{th}$ order (dot-dashed) and $10^{th}$ order (dotted)
Taylor expansion of $V_{LJ}$.
The points A and B are characterized by $V(x)= 1\%\ V(x_0)$, i.e.\ $99\%$
of the dissociation energy from the well bottom.
}
\end{figure}

To start, we set $N_{max}=2$, i.e.\ $V_d=0$ and analyze the pure Morse 
potential.
By imposing equal position of the minimum, equal well depth, and equal
value of the second derivative at the minimum, we fix all three Morse
parameters, and obtain the dashed curve drawn in Fig.~\ref{LJMorsebuca}.
These conditions are the natural choice to ensure an overall agreement,
which however is not particularly good, as the LJ and Morse potential are the
prototypes of two quite different behaviors.
As expected, the two potentials differ the most far from the equilibrium 
distance, in the approach to the dissociation limit.

\begin{table}
\centerline{
\begin{tabular}{lll}
\hline
    & $N_{max}=4$ & $N_{max}=6$\\
$\delta_{RMS}$   & 0.0046   &  0.00043\\
\hline
$\alpha$ & 0.10891 & 0.078138 \\
$V_0$    & 2.507 & 4.8698 \\
$a_3$    & 2.124 & 7.6676 \\
$a_4$    & 0.617 & 5.9625 \\
$a_5$    &       & 2.8206 \\
$a_6$    &       & 0.6560 \\  
\hline
\end{tabular}}
\caption{\label{FitLJParam}
Best fit parameters in the expansion (\ref{Morse-Exp}) with $N_{max}=4,6$
to the reference LJ potential with $\Sigma=31$, $A_{LJ}=4$ (so that the well
depth is unity).
}
\end{table}

We then use the full expansion (\ref{Morse-Exp}) truncated at $N_{max}=4$ and
$N_{max}=6$.
We impose the following conditions:
(i)
same location of the minimum (fix value of $x_0$),
(ii)
same values of second and third derivatives at the minimum (fix a relation
between $\alpha$ and $V_0$, and a relation involving $\alpha$ and $V_0$ and
$a_3$),
(iii)
same well depth (fix $V_0+\sum_i (-1)^ia_i$).
These 4 conditions reduce the number of adjustable parameters to $N_{max} - 3$.
To determine these residual free parameters, we minimize the RMS difference
of $V(x)$ and $V_M(x)+V_d(x)$ for approximately $10^3$ equally spaced
$x$-points between the points A, B of Fig.~\ref{LJMorsebuca}, where $V(x)=
1\%\ V(x_0)$, i.e.\ $99\%$ of the dissociation energy measured from the
well bottom.
The best fit parameters and the corresponding RMS discrepancy of the fit
are reported in Table~\ref{FitLJParam}.
Note that in both calculations the fitting potential is globally
well-behaved, as guaranteed by $a_{N_{Max}}>0$.
The agreement between the target potential $V_{LJ}$ and the approximating
expansion $V_M+V_d$ is very good already for $N_{max}=4$ (one fitted
parameter only), and the RMS discrepancy improves by about one order of
magnitude for $N_{max}=6$ (three fitted parameters).
Both fitted potentials are visibly indistinguishable from $V_{LJ}$, and
have therefore not been drawn in Fig.~\ref{LJMorsebuca}.
The fit is essentially independent of the density of the mesh points,
except for the choice of the cut at the repulsive side. The $99\%$ of the 
dissociation is a reasonable compromise to address the bound states.

\begin{table}[]
\begin{center}
\begin{tabular}{llll}
\hline
$n$ &	LJ (exact)  &	 $N_{max}=4$             &	 $N_{max}=6$ 	\\
\hline
0     & -0.88237 &	-0.88259   (-0.02\%) &	-0.88240   (-0.002\%)	\\
1     & -0.67488 &	-0.67592   (-0.10\%) &	-0.67494   (-0.006\%) 	\\
2     & -0.50142 &	-0.50387   (-0.25\%) &	-0.50146   (-0.004\%)	\\ 
3     & -0.35948 &	-0.36333   (-0.39\%) &	-0.35942   (0.005\%)	\\
4     & -0.24637 &	-0.25098   (-0.46\%) &	-0.24623   (0.014\%)	\\
5     & -0.15927 &	-0.16343   (-0.42\%) &	-0.15916   (0.011\%)	\\ 
6     & -0.09514 &	-0.09751   (-0.24\%) &  -0.09520   (-0.006\%)	\\
7     & -0.05078 &      -0.05040   (0.04\%) &	-0.05100   (-0.022\%)	\\
8     & -0.02278 &      -0.01973   (0.30\%)  &  -0.02286   (-0.008\%)	\\
9     & -0.00754 &      -0.00356   (0.40\%)  &  -0.00711   (0.043\%) 	\\
10    & -0.00126 &      (missing)            &  -0.00058   (0.068 \%)   \\
\hline
\end{tabular}
\end{center}
\caption{\label{LJMPMOGenRiep}
Bound-state energies for the LJ potential ($\Sigma=31,~A_{LJ}=4$) computed by 
numeric finite-differences solution of the Schr\"odinger equation (units of
$\hbar=\mu=1$), compared to the eigenvalues of the problem associated to
the best fit $V_M(x)+V_d(x)$, solved by diagonalization on the QNSB including
$N_s=40$ states. All reported figures are significant. 
In brackets the difference in percent of the well depth $E_d=A_{LJ}/4$.
}
\end{table}

This accurate fit of $V(x)$ produces extremely good accord for the
corresponding Schr\"odinger eigenvalues, as shown in
Table~\ref{LJMPMOGenRiep}.
A fairly accurate spectrum ($\Delta_{40}=0.46\% $) is obtained already for
$N_{max}=4$: the highest eigenvalue is missing but, due to its proximity to
dissociation, it is indeed very difficult to determine.
The fit with $N_{max}=6$ recovers the missing eigenstate, and improves 
substantially the accuracy of the eigenvalues ($\Delta_{40}=0.068\%$), 
determined precisely by the state closest to dissociation.
\begin{table}
\centerline{
\begin{tabular}{llllll}
\hline
$n$  & LJ        & Taylor 4           & Taylor 4            & Taylor 10          & Pure Morse          \\
     &   (exact) & (exact)            & (pert. theory)      &(exact)             &(analytical)         \\
\hline
0  & -0.8824    & -0.8791 (0.3\%)   & -0.8842 (-0.2\%)   & -0.8824 (0.001\%) & -0.8818   (0.06\%) \\ 
1  & -0.6749    & -0.6419 (3\%)     & -0.6899 (-1.5\%)   & -0.6742 (0.07\%)  & -0.6677   (0.7\%)  \\ 
2  & -0.5014    & -0.3868 (11\%)    & -0.5588 (-6\%)     & -0.4913 (1\%)     & -0.4833   (1.8\%)  \\ 
3  & -0.3595    & -0.1094 (25\%)    & -0.5079 (-15\%)    & -0.3065 (5\%)     & -0.32862  (3\%)    \\ 
4  & -0.2464    &  (positive & -0.5542 (-30\%)    & -0.0989 (15\%)    & -0.20369  (4\%)    \\ 
5  & -0.1593    &   energy                  & -0.7149 (-55\%)    &  (positive & -0.10850  (5\%)    \\ 
6  & -0.0951    &    $\dots$)                &  $\dots$            &          energy           & -0.04304  (5\%)    \\ 
7  & -0.0508    &                    &                     &         $\dots$)   & -0.00732  (4\%)    \\ 
8  & -0.0228    &                    &                     &                    & (missing)           \\ 
9  & -0.0075    &                    &                     &                    & (missing)           \\ 
10 & -0.0013    &                    &                     &                    & (missing)           \\ 
\hline
\end{tabular}
}
\caption{\label{LJRiepTayMor}
Comparison of the exact $V_{LJ}$ eigenstates with those of the $4^{th}$ order
traditional Taylor expansion (dot-dashed curve in Fig. \ref{LJMorsebuca}) 
obtained by exact integration of the differential problem, and by $2^{nd}$ 
order perturbation theory. The eigenvalues of the pure Morse curve (dashed in
Fig. \ref{LJMorsebuca}) are also reported for comparison. The differences in 
brackets are in percent of the well depth $E_d$.
}
\end{table}
Note that the absolute accuracy of the eigenstates is very close to 
the accuracy of the underlying functional fit of the potential.
 
For comparison with the traditional approach, Table \ref{LJRiepTayMor}
lists the exact LJ eigenenergies aside with (i) the approximate spectra
computed for the relevant Taylor expansion
$V(x)=\tilde{V}_0~+~\sum_{j=2}^{\tilde{N}}~c_j~(x-x_0)^j$ truncated at
$\tilde{N}=4$ and $10$ (drawn in Fig.~\ref{LJMorsebuca}); (ii) the spectrum
obtained by standard second-order perturbation theory applied to the
harmonic oscillator, based on the fourth-order Taylor expansion; (iii) the
spectrum of the pure Morse term drawn in Fig.~\ref{LJMorsebuca}.
It is apparent that the traditional approach can only reach the accuracy of
a few percent on few low-lying levels, but fails completely starting
already at moderately low overtones.  In contrast, even the rough
approximation of the pure Morse term provides level estimates within few
percent of the $V_{LJ}$ levels throughout the spectrum.
The expanded potential yields a good global description of the spectrum.
It would be straightforward to apply the present method to the recently
determined adiabatic potential of the Ar$_2$ dimer \cite{Patkowski05}, but
we prefer to consider in detail a molecule where PES is less close to the
LJ function.

\subsection{The H$_2$ molecule}

\begin{table}
\centerline{
\begin{tabular}{lllll}
\hline
                         & Morse            & $N_{max}=4$        & $N_{max}=12$          \\
$\delta_{RMS}$ [Ha]      & 0.035            & 0.006              & 0.00003            \\
$\delta_{RMS}$ [cm$^{-1}$] & 7760           & 1320               & 5.5                \\
$\delta_{RMS}/E_d$       & 0.20             & 0.03               & 0.0001             \\
\hline
$\alpha$                 & 1.318 $a_0^{-1}$  & 1.223 $a_0^{-1}$  & 0.840657 $a_0^{-1}$\\
$V_0$                    & 0.17447498 Ha     & 0.120  Ha         & 0.26147   Ha       \\
$a_3$                    &                   & -0.030 Ha         & 0.09233   Ha       \\
$a_4$                    &                   & 0.024  Ha         & 0.06147   Ha       \\
$a_5$                    &                   &                   & 0.03643   Ha       \\
$a_6$                    &                   &                   & -0.00995  Ha       \\
$a_7$                    &                   &                   & -0.02819  Ha       \\
$a_8$                    &                   &                   & 0.04013   Ha       \\
$a_9$                    &                   &                   & 0.02555   Ha       \\
$a_{10}$                 &                   &                   & -0.04034  Ha       \\
$a_{11}$                 &                   &                   & 0.01243   Ha       \\
$a_{12}$                 &                   &                   & 0.00025   Ha       \\
\hline
\end{tabular}
}
\caption{
Fit parameters for model potential (\ref{Morse-Exp}),
$N_{max}=2,4,12$.\label{FitH2Param}
}
\end{table}

\begin{figure}
\centerline{\epsfig{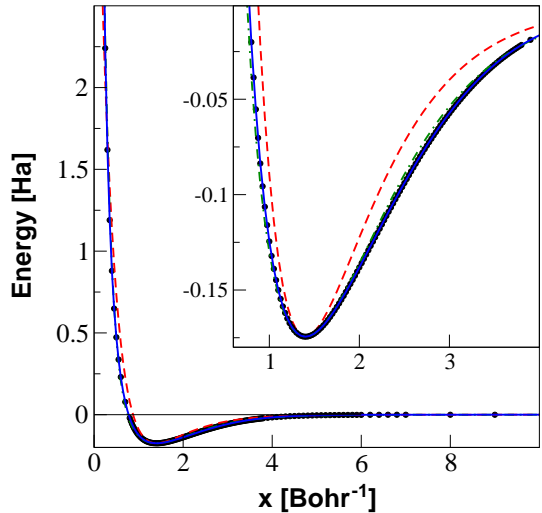}}
\caption{ \label{H2-fig}
Adiabatic potential for H$_2$, from Ref.~\cite{SLR87} (dots).
Best fit Morse potential with $\alpha$ as only adjusted parameter (dashed),
and fit function (\ref{Morse-Exp}), with $N_{max}=4$ (dot-dashed) and
$N_{max}=12$ (solid). Inset: detail of the minimum region, where the small
inaccuracy of the dot-dashed curve makes it distinguishable from the solid
curve.
}
\end{figure}

We acquire the accurate {\it ab-initio} determination of the H$_2$
adiabatic  potential of Refs.~\cite{SLR87}.
In our parameterization, we choose to fix both the well depth to the very
highly accurate value of $E_d=0.17447498$ Ha, and the minimum position
$x_0=1.4011\,a_0$ \cite{KolWol2}.
The other $N_{max}-1$ parameters, including a constant setting the overall 
energy zero,
are all adjusted by minimization of the RMS deviation $\delta_{RMS}$ 
from the {\it ab-initio} points provided in Ref.~\cite{SLR87}.
The mesh consists of all 169 available points ranging from $0.2\,a_0$ to 
$18\,a_0$ representing the effective BO potential for H$_2$ as computed in  
Ref.~\cite{SLR87}. The fits include all available points. In order to enhance 
the convergence in the bound-state energy range, the 
points above $1\%$ $V(x_0)$ are weigthed $1/9$ of those in the well region. 
A first fit with the Morse potential term alone yields a rather poor
approximation of $V_{H_2}$ ($\delta_{RMS}=0.035$ Ha), as shown in
Fig. \ref{H2-fig}.
We then fit the expansion (\ref{Morse-Exp}) with $N_{max}=4$ and $12$,
obtaining $\delta_{RMS}=0.006$ and $0.00003$ Ha respectively.
The best fit parameters are listed in Table \ref{FitH2Param}.
Figure~\ref{H2-fig} reports the resulting profiles.

\begin{table}
\centerline{
\begin{tabular}{llllll}%\label{EigH2}                      60                        40
\hline
$n$ &  Ref.~\cite{SLR87}  & Spline        & Pure Morse       & Eq.~(\ref{Morse-Exp}) $N_{max}=4$  & Eq.~(\ref{Morse-Exp}) $N_{max}=12$  \\ 
  & cm$^{-1}$            & cm$^{-1}$     &  cm$^{-1}$       & cm$^{-1}$                          & cm$^{-1}$          \\ 
\hline
0 & -36118               & -36112        & -35524  (2\%)    & -36136   (-0.06\%)                 & -36113 (-0.002\%)  \\ 
1 & -31957               & -31948        & -30298  (4\%)    & -31963   (-0.04\%)                 & -31948 (-)\\ 
2 & -28031               & -28021        & -25488  (7\%)    & -27979   (0.1\%)                   & -28020 (0.001\%)   \\ 
3 & -24336               & -24324        & -21093  (8\%)    & -24203   (0.3\%)                   & -24324 (-) \\ 
4 & -20868               & -20854        & -17114  (10\%)   & -20651   (0.5\%)                   & -20856 (-0.004\%)  \\ 
5 & -17626               & -17612        & -13550  (11\%)   & -17340   (0.7\%)                   & -17614 (-0.005\%)  \\ 
6 & -14612               & -14598        & -10402  (11\%)   & -14282   (0.8\%)                   & -14599 (-0.004\%)  \\ 
7 & -11830               & -11815        & -7670   (11\%)   & -11491   (0.8\%)                   & -11815 (-) \\
8 & -9287                & -9272         & -5353   (11\%)   & -8979    (0.8\%)                   & -9271  (0.003\%)   \\ 
9 & -6994                & -6980         & -3451   (9\%)    & -6755    (0.6\%)                   & -6979  (0.003\%)   \\ 
10& -4967                & -4955         & -1965   (8\%)    & -4829    (0.3\%)                   & -4955  (-) \\ 
11& -3230                & -3220         & -895    (6\%)    & -3213    (0.02\%)                  & -3222  (-0.006\%)  \\ 
12& -1816                & -1807         & -240    (4\%)    & -1915    (-0.3\%)                  & -1810  (-0.01\%)   \\
13& -766                 & -760          & -1.     (2\%)    & -943     (-0.5\%)                  & -761   (-0.002\%)  \\ 
14& -145                 & -142          &                  & -308     (-0.4\%)                  & -135   (0.02\%)    \\ 
15&                      &               &                  & -7                                 &                    \\
\hline
\end{tabular}
}
\caption{\label{EigH2}
Comparison of the computed H$_2$ bound-state energies: Ref.~\cite{SLR87} 
includes nonadiabatic corrections; when these corrections are left out, 
and all bound states of the splined potential of Ref.~\cite{SLR87} are 
computed by numerical solution of the differential Schr\"{o}dinger problem, 
the energies of the second column are obtained. The successive columns 
report eigenvalues of the pure Morse term and of the expansion 
(\ref{Morse-Exp}), fitted to the same adiabatic potential of Ref.~\cite{SLR87}. 
Discrepancies to the energies in the second column are reported in brackets: 
those of less than $0.001\%$ are replaced by a slash.
}
\end{table}

Complete exclusion of the points above $1\%$ $V(x_0)$ produces better 
agreement of low-order fits, but makes higher-order fits ($N_{max}>8$) 
unstable.
On the other hand, enforcing equal accuracy to all points (no weights) makes 
all fits much less accurate.

The model Hamiltonian is solved on a QNSB including the first $N_s= 60$ 
states:
the computed levels are summarized in Table~\ref{EigH2}.
The obtained eigenenergies are to be compared with those obtained by
numerical integration of the Schr\"{o}dinger problem of a third-order
spline going through the points of Ref.~\cite{SLR87}. These reference
energies (second column of Table \ref{EigH2}) differ from the levels
obtained in Ref.~\cite{SLR87} in non including the nonadiabatic
corrections.  The agreement improves systematically as $N_{max}$ increases,
with a worst discrepancy of 324 cm$^{-1}$ for $N_{max}=4$, which reduces to
7 cm$^{-1}$ for $N_{max}=12$. If the state closest to dissociation is not
considered, all other levels are computed within 2.4 cm$^{-1}$ of their
reference values.

In fact the $N_{max}=12$ spectrum could equally well be computed on a QNSB
of $N_s=30$ states only, and the same cm$^{-1}$ figures are obtained. The
reason is that the underlying Morse term is so wide and deep that it could
host $26$ bound states, so that even the pure Morse basis would be almost
complete for the bound states of the well at hand.
The $N_{max}=4$ calculation generates an extra spurious bound state whose 
precise energy location converges only very slowly with $N_s$, while all other 
states require only $N_s=60$ to remain stable to cm$^{-1}$.
This is due to the narrow shallow Morse well underlying this expansion.
In such a kind of expansion, the QNSB is absolutely necessary: the Morse 
basis would produce very poorly converged spectra.

\section{Discussion and conclusions}
\label{discussion:sec}

In this work we propose a technique to compute molecular vibrational
spectra including anharmonic effects using an appropriate virtually
complete Hilbert-space basis, the QNSB (\ref{QNSB}), and a smart
prescription for the form of the model potential, namely
Eq.~(\ref{Morse-Exp}).
The substantial advantage of the proposed combination of Hamiltonian and basis 
resides in the fully algebraic form of the matrix elements, which can be 
computed quickly and efficiently, and in the sparse form of the final matrix. 
The model potential of Eq.~(\ref{Morse-Exp}) is sufficiently flexible to 
realize a good compromise between 
local and global accuracy, with relatively low order expansions.
In principle, the virtually complete QNSB allows to determine accurate
vibrational levels in the whole bound-states energy range, and even to study
dissociation phenomena \cite{su11,Ben-Mol-Al}.

In the past, a similar approach was applied to restricted forms of the
expansion (\ref{Morse-Exp}).
For example, the PMO prescription \cite{ Huffak_1976_1,Huffak_1982} exclude
the $[\hat v(x)]^3$ term and determine the Morse parameters $\alpha$ and
$V_0$ in order to satisfy the conditions that the second and third
derivatives of the two potentials coincide at the minimum.
This fixes rigidly the global shape of all other terms, making that
expansion much less flexible and accurate.
It is eventually still possible to recover a good accuracy in that
approximation, but it is necessary to extend the expansion
(\ref{Morse-Exp}) to much higher order.
In fact, there is no special reason to fix the pure Morse contribution to
some local property of the addressed potential, such as derivatives at the
minimum, also because these local properties could even not be available in
the realistic case of a microscopic potential derived by {\it ab initio}
numerical calculations.

Other algebraic methods \cite{Iac-Lev_book,VanRoos-PhD,Oss-rev,Iac-Lev_I}
do not rely on an explicit potential, and this makes the calculation of
matrix elements of arbitrary operators and roto-vibrational couplings
impossible to carry out without the introduction of additional ad hoc
parameters.
The present method can instead generate all matrix elements of
$\hat{y}^i=e^{-\alpha i \hat{x}}$, of $\hat{x}^i e^{-\alpha'\hat x}$, and of
$\hat{p}_x^i$ by straightforward algebraic manipulation.

The method presented here is not the only one to involve smart choices of
expansion variables.
For example, in the direct-potential-fit methods
\cite{Ogilvie94,Ogilvie96,Coxon01,Ogilvie01,Coxon03,Coxon04,LeRoy04} the
roto-vibrational spectroscopy of diatomic molecules is addressed by taking
the experimental spectra as input.
Nonadiabatic effects can also be included.
The validity of the potential parameterization generated by such methods is
usually restricted to the energy range where data are available.
The method of the present work generates instead purely vibrational spectra
based on first-principle determination of the adiabatic PES, thus with
relatively constant accuracy through the whole bound-state energy range.
The PES parameterization proposed here is based on {\it ab-initio}
electronic-structure calculations.  It would be difficult (and of little
interest) to attempt to use the present formalism to fit experimental
lines, as in the direct-potential-fit approach.
An extension to address nonadiabatic and rovibrational effects could
be developed in future work.

When this method is extended to a polyatomic context, the total basis size
and its sparseness are the main computational issues to consider.
If in the present scheme spectroscopic accuracy can be obtained with as few
as $30$ states per degree of freedom, the total basis size could still
approach $30^6\simeq 10^9$ states for a tetraatomic molecule.
This size could exceed the computational power available, even if the
matrix to diagonalize is sparse.
As common practice in quantum chemical calculations, a truncated basis
could be selected by accepting slightly less accurate spectra.
This truncation may target specific 1D oscillators: in this case, when
$N_s$ is reduced to the number $[s]+1$ of the underlying Morse term, the
QNSB and the Morse basis are equivalent and yield equivalent spectra.
If further reduction is needed, the convergency of the two resulting
spectra shows a characteristically different deterioration.
Truncation of the Morse basis affects greatly the high-energy states near
dissociation: these disappear rapidly one by one, without significant
modification of the low-energy levels. On the contrary, the reduction of
QNSB size affects eigenenergies throughout the whole spectrum, but with a
much slower decrease of the number of bound states.
Accordingly, the two basis can be both used with success, keeping 
in mind their different behavior in different situations.
Alternatively, standard truncation of the full many-oscillator basis, based
on the relative perturbative weight of the states \cite{DelMonte05,Handy04}
is also made efficient by the very sparse structure of the matrix elements.

Extension of the method presented here to vibrational spectra of polyatomic
molecules is not especially more complicated than the formally similar
traditional harmonic-oscillator based expansion.
In particular it lends itself naturally to both diagonalization (Lanczos /
Davidson) \cite{Wyatt98,Pochert00,Callegari03,Gruebele96} and
Green-function--based \cite{Stuchebrukhov93,DelMonte05} iterative methods
of solution.
Special care must be put in making sure that the total polyatomic 
adiabatic potential surface (e.g.\ fitted on {\it ab initio} data) 
has a single absolute minimum.
This global property is easier to implement than on traditional 
power-series methods where the global shape of the approximate 
potential is dictated by local properties at the equilibrium
geometry.
A separate treatment of ``soft'' torsional coordinates may be needed,
exactly like within the traditional normal-coordinate harmonic-oscillator
based schemes.
Once the extension to the polyatomic case is realized, the present method
will compete successfully with the traditional methods, to yield {\it
ab-initio} vibrational spectra of acceptable accuracy up to the
high-overtones near-dissociation energy range.

\appendix 

\section{Matrix elements on the QNSB}\label{Matel6expl}

We report here the nonzero matrix elements of the cubic term
$[\hat v(x)]^3$:
\begin{equation}
\langle \phi_n|[\hat v(x)]^3| \phi_{n-3}\rangle = -\frac{C_{n -2 } C_{n -1} C_{n}}{(1 + 2 s)^3} 
\end{equation}
\begin{equation}
\langle \phi_n|[\hat v(x)]^3| \phi_{n-2}\rangle = \frac{3C_{n -1}C_{n}(2 n - 2 [s] -3)}{(1 + 2s)^3}
\end{equation}
\begin{eqnarray}
 &\langle \phi_n|[\hat v(x)]^3| \phi_{n-1}\rangle = -\frac{C_{n}}{(1 + 2s)^3}(12n^2 - 24n + 13 + C_{n -1}^2 + C_{n}^2 + C_{n + 1}^2 \\ \nonumber
&+ 12[s]^2 - 24n[s])
\end{eqnarray}
\begin{eqnarray}
 &\langle \phi_n|[\hat v(x)]^3| \phi_{n}\rangle = \frac{1}{(1 + 2s)^3}[C_{n}^2(6n - 6[s] -5) + C_{n + 1}^2(6n - 6[s] - 1) \\ \nonumber
&+ (2n - 2[s] -1)^3]
.
\end{eqnarray}

Similarly, the nonzero matrix elements of the quartic term $[\hat v(x)]^4$ read:
\begin{equation}
\langle \phi_n|[\hat v(x)]^4| \phi_{n-4}\rangle = \frac{C_{n -3 } C_{n -2 } C_{n -1} C_{n}}{(1 + 2 s)^4} 
\end{equation}
\begin{equation}
 \langle \phi_n|[\hat v(x)]^4| \phi_{n-3}\rangle =\frac{8 C_{n -2 } C_{n -1} C_{n} (2 - n +  [s])}{(1 + 2 s)^4 }
\end{equation}
\begin{eqnarray}
\hspace{-1.4cm}& \langle \phi_n|[\hat v(x)]^4| \phi_{n-2}\rangle =  \frac{C_{n -1} C_{n}}{(1 + 2 s)^4} (58 - 72 n + 24 n^2 + C_{n -2}^2 + C_{n -1}^2 + C_{n}^2 \\ \nonumber
&+ C_{n + 1}^2 + 72  [s] - 
    48 n  [s] + 24  [s]^2)
\end{eqnarray}

\begin{eqnarray}
\hspace{-1cm}&\langle \phi_n|[\hat v(x)]^4| \phi_{n-1}\rangle =
- \frac{4 C_{n}}{(1 + 2 s)^4}  [-10 + 26 n - 24 n^2 + 8 n^3 - C_{n+ 1}^2 \\ \nonumber
&+ 2 n C_{n + 1}^2 + 
    C_{n -1}^2 (2 n -3 - 2 [s]) 
+ 2 C_{n}^2 (n -1 -  [s]) - 26  [s] 
+    48 n  [s] \\ \nonumber
& - 24 n^2  [s] - 2 C_{n + 1 }^2  [s] - 24  [s]^2 + 
    24 n  [s]^2 - 8  [s]^3] 
\end{eqnarray}

\begin{eqnarray}
&\langle \phi_n|[\hat v(x)]^4| \phi_{n}\rangle =  \frac{1}{(1 + 2 s)^4}  \{C_{n}^4 + C_{n +1}^4 + (1 - 2n + 2[s])^4 + 
 C_{n+1}^2 \times  \\ \nonumber 
&(2- 8n + 24n^2 + C_{n +2}^2  +(8 - 48n)[s] + 
   24[s]^2) + C_{n}^2 [C_{n - 1}^2 \\ \nonumber
&+ 2(9 - 20n + 12 n^2 + C_{n +1}^2 + 20[s] - 24 n [s] + 12 [s]^2)]\}
.
\end{eqnarray}

The nonzero matrix elements of the  term $[\hat v(x)]^5$ read:
\begin{equation}
\langle \phi_n|[\hat v(x)]^5| \phi_{n-5}\rangle = -\frac{C_{n -4} C_{n -3} C_{n -2} C_{n -1} C_{n}}{(1 + 2 s)^5} 
\end{equation}

\begin{equation}
\langle \phi_n|[\hat v(x)]^5| \phi_{n-4}\rangle = -\frac{5 C_{n -3} C_{n -2} C_{n -1} C_{n}}{(1 + 2 s)^5} (2n -5  - 2[s])
\end{equation}

\begin{eqnarray}
\hspace{-1.4cm}& \langle \phi_n|[\hat v(x)]^5| \phi_{n-3}\rangle =  -\frac{C_{n -2} C_{n -1} C_{n}}{(1 + 2 s)^5}  (170 - 160 n + 40 n^2 + C_{n -3}^2 \\ \nonumber
&+ 
  C_{n -2}^2 + C_{n -1}^2 + C_{n}^2 + C_{n + 1}^2 + 160 [s] - 
  80 n [s] + 40 [s]^2)
\end{eqnarray}

\begin{eqnarray}
\hspace{-1.4cm}& \langle \phi_n|[\hat v(x)]^5| \phi_{n-2}\rangle =  -\frac{C_{n -1} C_{n}}{(1 + 2 s)^5} [330 - 580 n + 360 n^2 - 80 n^3 + 13 C_{n}^2 \\ \nonumber
&-  10 n C_{n}^2 + 9 C_{n+ 1}^2 - 10 n C_{n + 1 }^2 + 580 [s] - 720 n [s] + 240 n^2 [s] + 10 C_{n}^2 [s] \\ \nonumber
&+ 10 C_{n + 1}^2 [s] + 360 [s]^2 - 240 n [s]^2 + 80 [s]^3 + C_{n -1}^2 (17 - 10 n + 10 [s]) \\ \nonumber
&+ C_{n -2}^2 (21 - 10 n + 10 [s])]
\end{eqnarray}

\begin{eqnarray}
\hspace{-1.5cm}& \langle \phi_n|[\hat v(x)]^5| \phi_{n-1}\rangle =  -\frac{C_{n}}{(1 + 2 s)^5} [121 - 400 n + 520 n^2 - 320 n^3 + 80 n^4 + C_{n -1}^4 \\ \nonumber
&+ C_{n}^4 + 
  14 C_{n + 1}^2 - 40 n C_{n + 1}^2 + 40 n^2 C_{n + 1}^2 + C_{n + 1}^4 + 
  C_{n + 1}^2 C_{n + 2}^2 + 400 [s] \\ \nonumber
&- 1040 n [s] + 
  960 n^2 [s] - 320 n^3 [s] + 
  40 C_{n + 1}^2 [s] - 80 n C_{n + 1}^2 [s] + 
  520 [s]^2 \\ \nonumber
&- 960 n [s]^2 + 480 n^2 [s]^2 + 
  40 C_{n + 1}^2 [s]^2 + 320 [s]^3 - 
  320 n [s]^3 + 80 [s]^4 \\ \nonumber
&+ 2 C_{n}^2 (21 - 40 n + 20 n^2 + C_{n + 1}^2 - 40 (n -1) [s] + 
    20 [s]^2) + C_{n -1}^2 (94 \\ \nonumber
&- 120 n + 40 n^2 + C_{n -2}^2 +    2 C_{n}^2 
+ C_{n + 1}^2 + 120 [s] - 80 n [s] + 
					 40 [s]^2)]
\end{eqnarray}
\begin{eqnarray}
\hspace{-1.5cm}& \langle \phi_n|[\hat v(x)]^5| \phi_{n}\rangle =  -\frac{1}{(1 + 2 s)^5} \{ C_{n + 1}^4 (1 - 10 n + 10 [s]) 
 \\ \nonumber
&+ C_{n}^4 (9- 10 n + 10 [s]) + 
 C_{n + 1}^2 [2 - 20 n + 40 n^2 - 80 n^3 + 20 (1 - 4 n \\ \nonumber
&+ 12 n^2) 
    [s] - 40 (6 n -1) [s]^2 + 80 [s]^3 + 
   C_{n + 2}^2 (10 [s] -3 - 10 n)] \\ \nonumber
&+ 
 C_{n}^2 \{C_{n -1}^2 (13 - 10 n + 10 [s]) + 2 [29 - 90 n + 100 n^2 - 40 n^3 \\ \nonumber
&+ 10 (9 - 20 n + 12 n^2) [s] - 
     20 (6 n -5) [s]^2 + 40 [s]^3 \\ \nonumber
&+ 
      C_{n + 1}^2 (5 - 10 n + 10 [s])]\} + (1 - 2 n + 2 [s])^5\}
.
\end{eqnarray}

The nonzero matrix elements of the term  $[\hat v(x)]^6$ read:

\begin{equation}
\langle \phi_n | [\hat v(x)]^6| \phi_{n-6} \rangle=\frac{C_{n -5 } C_{n -4} C_{n -3} C_{n -2} C_{n -1} C_{n}}{(1+2s)^6}
\end{equation}

\begin{equation}
\langle \phi_n | [\hat v(x)]^6| \phi_{n-5} \rangle=\frac{12 C_{n -4} C_{n -3} C_{n -2} C_{n -1} C_{n}}{(1+2s)^6} (3 - n +  [s])
\end{equation}

\begin{eqnarray}
&\langle \phi_n | [\hat v(x)]^6| \phi_{n-4} \rangle= \frac{C_{n -3} C_{n -2} C_{n -1} C_{n}}{(1+2s)^6} (395 - 300 n + 60 n^2 + C_{n -4}^2 \\ \nonumber
&+ C_{n -3}^2  
    +C_{n -2}^2 + C_{n -1}^2 + C_{n}^2 + C_{n + 1}^2 + 300  [s] - 120 n  [s] + 
    60  [s]^2)
\end{eqnarray}

\begin{eqnarray}
&\langle \phi_n | [\hat v(x)]^6| \phi_{n-3} \rangle= -\frac{4 C_{n -2} C_{n -1} C_{n} }{(1+2s)^6} [-380 + 510 n - 240 n^2 + 40 n^3  \\ \nonumber
&- 6 C_{n -1}^2 + 3 n C_{n -1}^2 - 5 C_{n}^2 + 
    3 n C_{n}^2 - 4 C_{n + 1}^2 + 3 n C_{n + 1}^2 + C_{n -3}^2 \times\\ \nonumber
&(-8 + 3 n - 3  [s]) + 
    C_{n -2}^2 (-7 + 3 n - 3  [s]) - 510  [s] + 480 n  [s] \\ \nonumber
&- 120 n^2  [s] - 3 C_{n -1}^2  [s] - 3 C_{n}^2  [s] - 
    3 C_{n + 1}^2  [s] - 240  [s]^2 \\ \nonumber
&+ 120 n  [s]^2 - 40  [s]^3]
   %\delta_{m,n  -3}
\end{eqnarray}

\begin{eqnarray}
&\langle \phi_n | [\hat v(x)]^6| \phi_{n-2} \rangle=  \frac{C_{n -1} C_{n}}{(1+2s)^6}  [1771 - 3960 n + 3480 n^2 - 1440 n^3 + 240 n^4 \\ \nonumber
&+ C_{n -2}^4 + C_{n -1}^4 + 107 C_{n}^2 - 
    156 n C_{n}^2 + 60 n^2 C_{n}^2 + C_{n}^4 + 59 C_{n + 1}^2 - 108 n C_{n + 1}^2 \\ \nonumber
&+ 60 n^2 C_{n + 1}^2 + 
    2 C_{n}^2 C_{n + 1}^2 + C_{n + 1}^4 + C_{n + 1}^2 C_{n + 2}^2 + 3960  [s] - 6960 n  [s] \\ \nonumber
&+ 4320 n^2  [s] - 960 n^3  [s] + 156 C_{n}^2  [s] - 120 n C_{n}^2  [s] + 
    108 C_{n + 1}^2  [s] - 120 n C_{n + 1}^2  [s] \\ \nonumber
&+ 3480  [s]^2 - 
    4320 n  [s]^2 + 1440 n^2  [s]^2 + 60 C_{n}^2  [s]^2 + 
    60 C_{n + 1}^2  [s]^2 + 1440  [s]^3 \\ \nonumber
&- 960 n  [s]^3 + 240  [s]^4 + 
    C_{n -1}^2 (179 - 204 n + 60 n^2 + 2 C_{n}^2 + C_{n + 1}^2 + 204  [s] \\ \nonumber
&- 120 n  [s] + 
      60  [s]^2) + C_{n -2}^2 (275 - 252 n + 60 n^2 + C_{n -3}^2 + 2 C_{n -1}^2 + C_{n}^2 \\ \nonumber
&+ 
      C_{n + 1}^2 + 252  [s] - 120 n  [s] + 60  [s]^2)] %\delta_{m,n -2}
\end{eqnarray}

\begin{eqnarray}
&\langle \phi_n | [\hat v(x)]^6| \phi_{n-1} \rangle= -\frac{4 C_{n}}{(1+2s)^6} [-91 + 363 n - 600 n^2 + 520 n^3 - 240 n^4 + 48 n^5  \\ \nonumber
&- 11 C_{n + 1}^2 + 42 n C_{n + 1}^2 - 
    60 n^2 C_{n + 1}^2 + 40 n^3 C_{n + 1}^2 - C_{n + 1}^4 + 3 n C_{n + 1}^4 \\ \nonumber
&+ 3 n C_{n + 1}^2 C_{n + 2}^2 + 
    C_{n -1}^4 (-5 + 3 n - 3  [s]) + 3 C_{n}^4 (n -1 -  [s]) - 363  [s] \\ \nonumber
&+ 1200 n  [s] - 1560 n^2  [s] + 960 n^3  [s] - 240 n^4  [s] - 
    42 C_{n + 1}^2  [s] + 120 n C_{n + 1}^2  [s]\\ \nonumber
&- 120 n^2 C_{n + 1}^2  [s] - 
    3 C_{n + 1}^4  [s] - 3 C_{n + 1}^2 C_{n + 2}^2  [s] - 600  [s]^2 + 1560 n  [s]^2 - 1440 n^2  [s]^2 \\ \nonumber
&+ 480 n^3  [s]^2 - 
    60 C_{n + 1}^2  [s]^2 + 120 n C_{n + 1}^2  [s]^2 - 520  [s]^3 + 960 n  [s]^3\\ \nonumber
& - 480 n^2  [s]^3 - 40 C_{n + 1}^2  [s]^3 - 240  [s]^4 + 240 n  [s]^4 - 48  [s]^5 + C_{n -1}^2 \times\\ \nonumber
&[-153 + 282 n - 180 n^2 + 40 n^3 - 3 C_{n + 1}^2 + 3 n C_{n + 1}^2 + 2 C_{n}^2 (-4 + 3 n - 3  [s])\\ \nonumber
& + 3 C_{n -2}^2 (n -2 -  [s]) -  282  [s] + 360 n  [s] - 120 n^2  [s] - 3 C_{n + 1}^2  [s] - 180  [s]^2 + 120 n  [s]^2 \\ \nonumber
&- 40  [s]^3] + 2 C_{n}^2 (-23 + 63 n - 60 n^2 + 20 n^3 + C_{n + 1}^2 (-2 + 3 n - 3  [s])\\ \nonumber
& - 3 (21 - 40 n + 20 n^2)  [s] + 60 (n -1)  [s]^2 - 20  [s]^3)]
   %\delta_{m,n -1}
\end{eqnarray}

\begin{eqnarray}
&\langle \phi_n | [\hat v(x)]^6| \phi_{n} \rangle= \frac{1}{(1+2s)^6} \{C_{n}^6 + C_{n +1}^6 + (1 - 2 n + 2 [s])^6 + 
 C_{n +1}^4 \{2 C_{n +2}^2 \\ \nonumber
&+ 3[1 - 4 n+ 20 n^2 + (4 - 40 n) [s] + 20 [s]^2]\} + 
 C_{n}^4 [2 C_{n -1}^2 + 3 (17 - 36 n + 20 n^2 + C_{n + 1}^2 \\ \nonumber
&+ (36 - 40 n) [s]  + 20 [s]^2)] + C_{n}^2 \{179 - 696 n + 1080 n^2 - 800 n^3 + 240 n^4 + C_{n - 1}^4  \\ \nonumber
&+ 3 C_{n + 1}^4+ 696 [s] - 2160 n [s] + 2400 n^2 [s] - 960 n^3 [s] + 1080 [s]^2 - 2400 n [s]^2 \\ \nonumber
&+ 1440 n^2 [s]^2 + 800 [s]^3 - 960 n [s]^3 + 240 [s]^4 + C_{n - 1}^2 (107 - 156 n + 60 n^2 \\ \nonumber
&+ C_{n -2 }^2 + 2 C_{n + 1}^2 + 156 [s] - 120 n [s] + 60 [s]^2) + 2 C_{n + 1}^2 [19 - 60 n + 60 n^2 \\ \nonumber
&+ C_{n + 2}^2 - 60 (2 n - 1) [s] + 60 [s]^2]\} + C_{n + 1}^2 \{3 - 24 n + 120 n^2 - 160 n^3 + 240 n^4 \\ \nonumber
&+ C_{n + 2}^4 + 24 (1 - 10 n + 20 n^2 - 40 n^3) [s] + 120 (1 - 4 n + 12 n^2) [s]^2 - 160 (6 n - 1) [s]^3 \\ \nonumber
&+ 240 [s]^4 + C_{n + 2}^2 [11 + 36 n + 60 n^2 + C_{n + 3}^2 - 12 (3 + 10 n) [s] + 60 [s]^2]\}\}
.
\end{eqnarray}


\newpage
\begin{thebibliography}{}
 
\bibitem{Carter97}
Carter,  S.; Bowman, J.\ M.; Harding, L.\ B.  Spectrochim.\ Acta\ A 1997, 53, 1179. 
  
\bibitem{Wyatt98}
Wyatt,  R.  J.\ Chem.\ Phys.\ 1998, 109, 10732. 
 
\bibitem{Pochert00}
Pochert,  J.; Quack, M.; Stohner, J.; Willeke, M.  J.\ Chem.\ Phys. 2000, 113, 2719. 
 
\bibitem{Stuchebrukhov93}
Stuchebrukhov,  A.\ A.; Marcus, R.\ A.  J.\ Chem.\ Phys. 1993, 98, 6044. 
 
\bibitem{DelMonte05}
Del Monte,  A.; Manini, N.; Molinari, L.\ G.; Brivio, G.\ P.  Mol.\ Phys.\ 2005, 103, 689. 
 
\bibitem{Csaszar97}
Cs{\'a}sz{\'a}r,  A.\ G.; Mills, I.\ M.  Spectrochim.\ Acta A 1997, 53, 1101. 
 
\bibitem{Koput98}
Koput,  J.; Peterson, K.\ A.  Chem.\ Phys.\ Lett.\ 1998, 283, 139. 
 
\bibitem{Miani00}
Miani,  A.; H\"anninen, V.; Horn, M.; Halonen, L.  Mol.\ Phys.\ 2000, 98, 1737. 
 
\bibitem{Callegari03}
Callegari,  A.; Pearman, R.; Choi, S.; Engels, P.; Srivastava, H.; Gruebele, M.; Lehmann, K.\ K.; Scoles, G.  Mol.\ Phys. 2003, 101, 551. 
 
\bibitem{Jensen00}
Jensen,  P.  Mol.\ Phys.\ 2000, 98, 1253. 
 
\bibitem{Jensen88}
Jensen,  P.  J.\ Mol.\ Spectrosc.\ 1988, 128, 478. 
 
\bibitem{Lemus04}
Lemus,  R.  J.\ Mol.\ Spectrosc.\ 2004, 225, 73. 
 
\bibitem{Halonen88}
Halonen,  L.; Carrington Jr, T.\.  J.\ Chem.\ Phys.\ 1988, 88, 4171. 
 
\bibitem{HCAO1}
Child,  M.\ S.; Halonen, L.  Adv.\ Chem.\ Phys.\ 1984, 57, 1. 
 
\bibitem{Morse_1929}
Morse,  P.\ M.  Phys.\ Rev.\ 1929, 34, 57. 
 
\bibitem{Huffak_1976_1}
Huffaker,  J.\ N.  J.\ Chem.\ Phys.\ 1976, 64, 3175. 
 
\bibitem{Huffak_1982}
Huffaker,  J.\ N.; Tran, L.\ B.  J.\ Chem.\ Phys.\ 1982, 76, 3838. 
 
\bibitem{Makar_1990}
Makarewicz,  J.  J.\ Phys. \ B: \ At. \ Mol. \ Opt. \ Phys.\ 1990, 24, 383. 
 
\bibitem{Iac-Lev_book}
Iachello,  F.; Levine, R.\ D.  Algebraic Theory of Molecules, Oxford University Press, 1995, p. 161. 
 
\bibitem{VanRoos-PhD}
van Roosmalen,  O.\ S.  Algebraic Descriptions of Nuclear, Molecular Rotation-Vibration Spectra, PhD Thesis, Univ.\ Gr\"oningen, 1982, unpublished. 
 
\bibitem{Oss-rev}
Oss,  S.  Adv.\ Chem.\ Phys.\  XCIII, 455. 
 
\bibitem{Iac-Lev_I}
Iachello,  F.; Levine, R.\ D.  J.\ Chem.\ Phys.\ 1982, 77 , 3046. 
 
\bibitem{Sage1}
Sage,  M.\ L.  Chem.\ Phys.\ 1978, 35, 375. 
 
\bibitem{PerBer1}
P\'erez-Bernal,  F.; Martel, I.; Arias, J.\ M.; G\'omez-Camacho, J.  Phys.\ Rev.\ A 2001, 63, 052111. 
 
\bibitem{PerBer2}
P\'erez-Bernal,  F.; Martel, I.; Arias, J.\ M.; G\'omez-Camacho, J.  Phys.\ Rev.\ A 2003, 67, 052108. 
 
\bibitem{su11}
Lemus,  R.; Arias, J.\ M.; G\'omez-Camacho, J.  J.\ Phys.\ A : Math.\ Gen.\ 2004, 37, 1805. 
 
\bibitem{SusyRev}
Cooper,  F.; Khare, A.; Sukhatme, U.\ P.  Phys.\ Rep.\ 1995, 251, 268. 
 
\bibitem{DongLemus}
Dong,  S.\ H.; Lemus, R.  Int.\ J.\ Quantum Chem.\ 2002, 86, 265. 
 
\bibitem{Ben-Mol-Al}
Moln\'ar,  B.; F\"{o}ldi, P.; Benedict, M.\ G.; Bartha, F.  Europhys.\ Lett. 2003, 61 (4), 445. 
 
\bibitem{Tennyson82}
Tennyson,  J.; Sutcliffe, B.\ T.  J.\ Chem.\ Phys. 1982, 77, 4061. 
 
\bibitem{Tennyson04}
Tennyson,  J.; Kostin, M.\ A.; Barletta, P.; Harris, G.\ J.; Polyansky, O.\ L.; Ramanlal, J.; Zobov, N.\ F.  Comp.\ Phys.\ Comm.\ 2004, 163, 85. 
 
\bibitem{Patkowski05}
Patkowski,  K.; Murdachaew, G.; Fou, C.-M.; Szalewicz, K.  Mol.\ Phys. 2005, 103, 2031. 
 
\bibitem{SLR87}
Schwartz,  C.; Le Roy, R.\ J.  J.\ Mol.\ Spectrosc.\ 1987, 121, 420. 
 
\bibitem{KolWol2}
Kolos,  W.; Wolniewicz, L.  J.\ Chem.\ Phys.\ 1968, 49, 404. 
 
\bibitem{Ogilvie94}
Ogilvie,  J.\ F.  J.\ Phys.\ B: At.\ Mol.\ Opt.\ Phys. 1994, 27, 47. 
 
\bibitem{Ogilvie96}
Ogilvie,  J.\ F.  J.\ Mol.\ Spectrosc. 1996, 180, 193. 
 
\bibitem{Coxon01}
Coxon,  J.\ A.; Molski, M.  Phys.\ Chem.\ Comm.\ 2001, 20, 1. 
 
\bibitem{Ogilvie01}
Ogilvie,  J.\ F.  Chem.\ Phys.\ Lett.\ 2001, 348, 447. 
 
\bibitem{Coxon03}
Coxon,  J.\ A.; Molski, M.  Spectrochim.\ Acta A 2003, 59, 13. 
 
\bibitem{Coxon04}
Coxon,  J.\ A.; Molski, M.  J.\ Mol.\ Spectrosc. 2004, 223, 51. 
 
\bibitem{LeRoy04}
Le Roy,  R.\ J.  J.\ Mol.\ Spectrosc. 2004, 228, 92. 
 
\bibitem{Handy04}
Handy,  N.\ C.; Carter, S.  Mol.\ Phys.\ 2004, 102, 2201. 
 
\bibitem{Gruebele96}
Gruebele,  M.  J.\ Chem.\ Phys. 1996, 104, 2453. 
\end{thebibliography}
\end{document}